\documentclass[conference]{IEEEtran}
\IEEEoverridecommandlockouts
\usepackage{cite}
\usepackage{amsmath,amssymb,amsfonts}
\usepackage{algorithmic}
\usepackage{graphicx}
\usepackage{textcomp}
\usepackage[ruled]{algorithm2e}
\usepackage{multirow}

\def\BibTeX{{\rm B\kern-.05em{\sc i\kern-.025em b}\kern-.08em
    T\kern-.1667em\lower.7ex\hbox{E}\kern-.125emX}}
\begin{document}

\title{Interpretable Parallel Recurrent Neural Networks with Convolutional Attentions for Multi-Modality Activity Modeling}

\author{\IEEEauthorblockN{Kaixuan Chen\IEEEauthorrefmark{1}, Lina Yao\IEEEauthorrefmark{1}, Xianzhi Wang\IEEEauthorrefmark{1}, Dalin Zhang\IEEEauthorrefmark{1}, Tao Gu\IEEEauthorrefmark{2}, Zhiwen Yu\IEEEauthorrefmark{3} and Zheng Yang\IEEEauthorrefmark{4}}

\IEEEauthorblockA{\IEEEauthorrefmark{1}School of Computer Science and Engineering, UNSW Sydney, Australia}
\IEEEauthorblockA{\IEEEauthorrefmark{2}School of Information Technology, RMIT University, Australia}
\IEEEauthorblockA{\IEEEauthorrefmark{3}School of Computer Science, Northwestern Polytechnical University, China}
\IEEEauthorblockA{\IEEEauthorrefmark{4}School of Software, Tsinghua University China}}

\maketitle

\begin{abstract}
Multimodal features play a key role in wearable sensor based human activity recognition (HAR).
Selecting the most salient features adaptively is a promising way to maximize the effectiveness of multimodal sensor data.
In this regard, we propose a ``collect fully and select wisely" principle as well as an interpretable parallel recurrent model with convolutional attentions to improve the recognition performance.
We first collect modality features and the relations between each pair of features to generate activity frames, and then introduce an attention mechanism to select the most prominent regions from activity frames precisely. The selected frames not only maximize the utilization of valid features but also reduce the number of features to be computed effectively. We further analyze the accuracy and interpretability of the proposed model based on extensive experiments. The results show that our model achieves competitive performance on two benchmarked datasets and works well in real life scenarios.
\end{abstract}

\begin{IEEEkeywords}
HAR, attention, deep learning, wearable sensors
\end{IEEEkeywords}

\section{Introduction}

HAR plays a key role in several research fields. It has gained broad attention due to the increasing popularity of ubiquitous environments, especially in health care and surveillance domains \cite{yao2016learning}.


Despite a large number of sensor-based recognition solutions proposed over the decade, we discover several limitations.
First, there is still a lack of comprehensive model representation to sensor signals in a way that different activities can be distinguished in a more expressive and effective ways.
With the recent advances in deep neural networks and the notable performance achieved by these methods in the community of HAR \cite{guo2016wearable},
Convolutional Neural Network (CNN) appears to be a promising candidate for building such models.
However, while CNN does well in capturing spatial relationships of features, it focuses merely on the features covered by the convolutional kernels but overlooks the correlation among non-adjacent features \cite{zhang2018cascade}.
Considering that most of the data collected by wearable sensors such as accelerometers and gyroscopes are tri-axis, in this paper, we transform sensor signals into new activity frames which not only capture the relationships between each pair of tri-axis signals but also contain the relations between each pair of single signals. The experiments show that our new representation is far more discriminative than traditional representations.

Second, the demerits of interperson variability and interclass similarity can greatly reduce system performance \cite{bulling2014tutorial}.
Interperson variability comes from the fact that the same activity can be performed differently by different people, and interclass similarity results from the similarity in the behavior patterns of different activities like walking and running. Both the above issues require the classifier to be task dependent, i.e., it should automatically extract the salient information indicative of the true activity and ignore the interclass similarity. To this end, we propose an attention based model, which is directly related to the HAR tasks, to address the problems of interperson variability and interclass similarity. 

Attention is originally a concept in biology and psychology that implies focusing the power of noticing or thinking on something special to achieve better cognitive processes.
The attention mechanisms have several advantages, the first being task dependence. Intuitively, the motion of different body parts has varied contributions to different activities \cite{yacoob1998parameterized}. For example, jumping mostly involves legs while running is related to both arms and legs. More specifically, recognizing the patterns of walking depends more on the acceleration of legs while distinguishing sitting from lying would rely more on the orientation. 
In this paper, we separate the data related to each body part to different modals, namely acceleration, angular velocity and magnetism. The attention mechanisms ensure that the system only focuses on the most contributing data and ignores the irrelevant sensors or modalities.

The second advantage of the attention mechanisms is that it opens the black box of deep neural networks to a certain degree.
While the inner mechanisms of neural networks remain implicit, interpretable neural network is becoming another trend in the machine learning and data mining fields. Taking CNN for example, when CNN recognizes a dog from an image, we tend to explicitly know that one filter distinguishes the dog head and another filter identifies the dog paw.
Back to activity recognition, the attention model not only provides the specific body parts it focuses on but also highlights the most contributing sensors and modals to distinguish diverse activities. The salient sensor data can be inferred from the glimpse patch (to be detailed in Section~\ref{sec:Glimpse Network} ).

The third advantage is that it reduces the computational cost significantly.
Usually, the dimension of the features expands as we extract the full spatial relationships among sensors, and the cost increases with the increase of input data dimension.
Most existing models process the entire data every time, resulting in high computational cost.
Some works \cite{ooi2016image, lai2014multilinear} aim to limit the input dimension using techniques such as dimensionality reduction and feature selection. However, feature processing comes with information loss, leading to a new trade-off  problem between accuracy and cost.
Inspired by human attention, our proposed method focuses on only one small patch of the data each time and goes to the next patch when necessary.
This method considerably reduces computational cost as well as information loss. 

In this paper, we tackle the HAR problems by transforming wearable sensor data into activity frames and deploying an interpretable parallel recurrent model with convolutional attentions, including one attention based LSTM and one activity frame based LSTM, to recognize activities. The main contributions of this work are summarized as follows:

\begin{itemize}

\item We transform the tri-axis sensor data into activity frames to extract the full relationships between data pairs. This enables the CNN to cover all features without overlooking any relationships between data pairs. 
\item We propose a parallel recurrent model including one attention based LSTM and one activity frame based LSTM to recognize activities. Firstly, the system focuses on only a small patch of the activity frame that contains the most salient information to avoid unnecessary cost on less important areas, by leveraging the attention based LSTM and combining reinforcement learning. Secondly, we deploy a activity frame based LSTM to exploit spatial and temporal information in time-series signals and capture the dynamics of the sensor data.
\item We examine our model on two public benchmarked datasets, PAMAP2 \cite{reiss2012introducing} and MHEALTH \cite{banos2015design} and perform extensive comparison with other methods, as well re-examine our approach on a new dataset collected in the real world named MARS. The experimental results show that our proposed model consistently outperforms a series of baselines and state-of-the-arts. 

\end{itemize}

\section{Our Model}

To fully collect effective information and wisely select salient features, our model contains two parts: (a) feature extraction
to firstly transform wearable sensor data into 2-D matrices. (b) an interpretable parallel recurrent model with convolutional attentions including one attention based LSTM and one activity frame based LSTM for activity recognition. Our attention based LSTM simulates the procedures of human brains processing visual information within several glimpses. The other one is activity frame based LSTM. Owing to the facts that the activity recognition largely depends on temporal information and that activity frames naturally capture serial relations, activity frame based model is more suitable for our scenarios.

\begin{figure}[htbp] 
\centering
\includegraphics[width=3.2in]{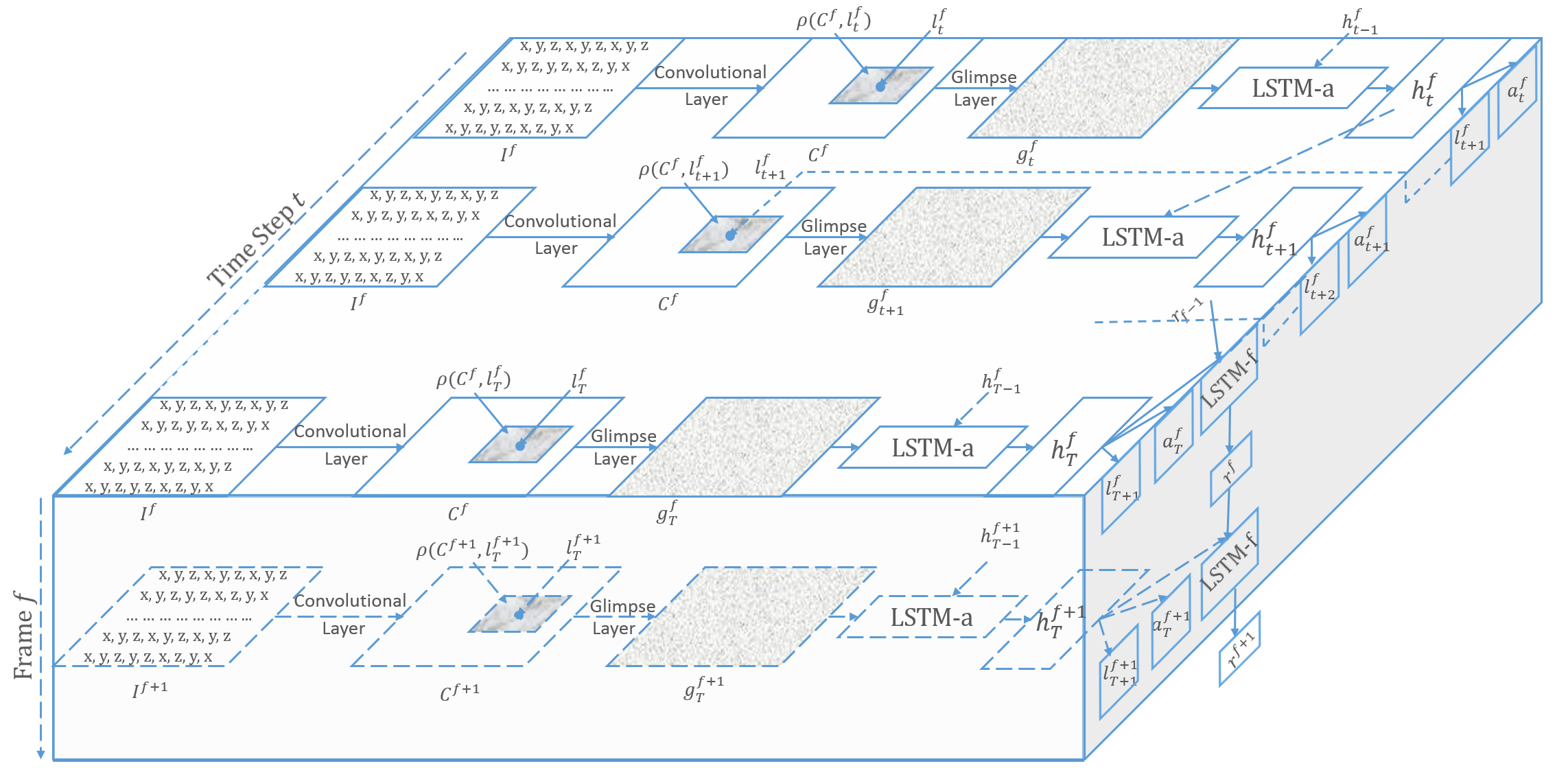} 
\caption{Work-flow of the Proposed Approach. Dashed arrows indicate the time step $t$ for attention based LSTM and the frame $f$ for activity frame based LSTM, respectively. For each time step $t$, the input frame goes through a convolutional network to obtain a higher-level representation $C^f$. We extract a retina region $\rho (C^f, l^f_t)$ at location $l^f_t$, which is decided by the last time step $t-1$. $\rho (C^f, l^f_t)$  next goes through a glimpse layer to get the glimpse $g^f_t$ as input of the attention based LSTM-$a$ which decides the action $a^f_t$ and the next location $l^f_{t+1}$. For the activity frame based LSTM-$f$ takes the last action of each frame $a_f^T$ as input and outputs the final prediction.}
\label{fig:model} 
\end{figure}

The above process is presented as a three-dimensional model in Figure~\ref{fig:model}, where the time step $t$ and frame $f$ represent the attention based LSTM and the activity frame based LSTM in our method, respectively.

\subsection{Input Representation}

As we transform the wearable sensor data into activity frames, the data are represented as three-dimensional vectors. Each sample $(\textbf{x, y})$ of the model consists of a 3-d vector $\textbf{x}$ and the activity label $\textbf{y}$.
Suppose  $X, Y, F$ denote activity frames' width, height, and number of frames, and $C$ represents the number of activity classes, we have:
$\textbf{x}\in R^{X\times Y\times F}$ and $\textbf{y}\in [1, ..., C]$.

\subsubsection{\textbf{Activity Frame}} 


There already exist some previous works that combine multimodal wearable sensor data for HAR in feature level. For example, Kunze et al. \cite{kunze2008dealing} concatenate acceleration and angular velocity into one vector and \cite{lara2012centinela,parkka2006activity,tapia2007real} combine acceleration and other modalities including microphone and GPS data. However, these works overlook the relations among sensors which are important to activity recognition. A popular method for extracting spatial relations is deep learning methods like CNN. Although CNN is proven to perform well in HAR \cite{jiang2015human, yang2015deep}, the accuracy is still not that satisfactory. In fact, CNN is originally proposed for images where each pixel is only related to its adjacent pixels and this small area can be easily covered by a kernel patch of a convolutional layer. However, it is still challenging to transform features to extract relations between each signal and the related signals for HAR. In many cases of HAR \cite{wang2017modeling}, the sensor data are arranged according to the physical connection of human body parts. For example, the sensor data of hands should be adjacent to the data of shoulders and the data of shoulders should be adjacent to the data of the waist, which should be followed by the data of hips, legs, and feet. Nevertheless, in the real world, activities always depend on more than one body part. For instance, running relies on the cooperation of arms and legs. In addition, the common Inertial Measurement Unit in wearable devices usually includes a tri-axis accelerometer, a tri-axis gyroscope, and a tri-axis magnetometer, and the degree to which these sensors contribute to different activities are various. This makes it even more important to find a representative transformation to extract the relationships between each pair of tri-axis sensor signals (e.g. acceleration and angular velocity) and each pair of single signals (e.g. the first dimension of acceleration and the second dimension of angular velocity). 

\begin{figure}[htbp] 
\centering
\includegraphics[width=3.4in]{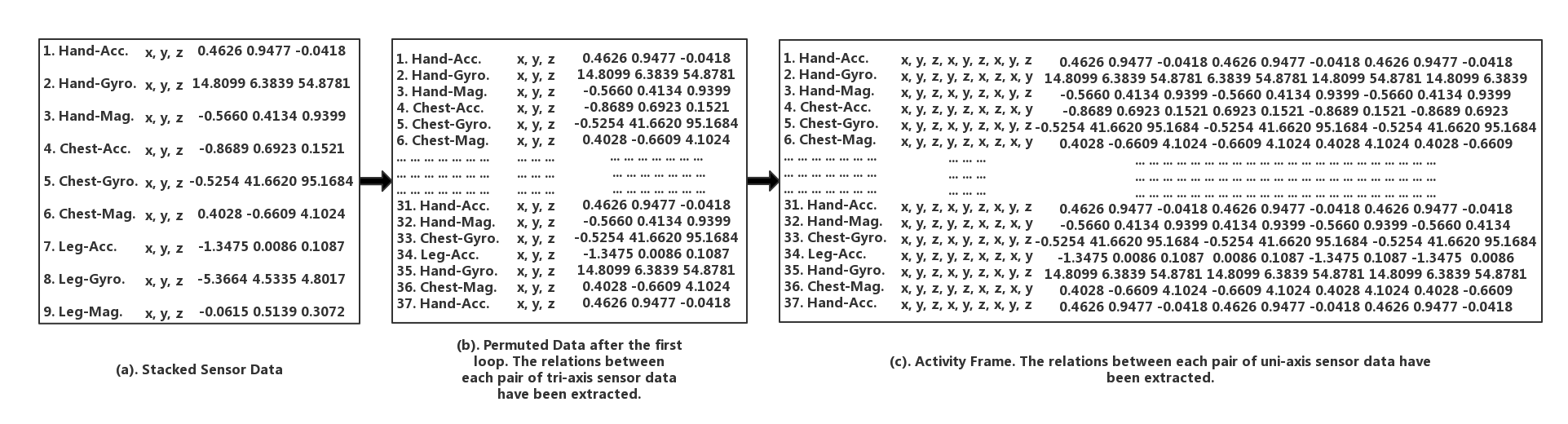}
  \caption{Transformation from sequences to frames}
  \label{fig:activity image}
\end{figure}

Figure~\ref{fig:activity image} shows the transformation process into activity frames. Each figure is comprised of four parts: sequence number, sensor location (hand, chest, leg) and modality (acceleration, angular velocity...), notations (x, y, z), and real data examples.
Algorithm~\ref{alg:activity frames} further illustrates the transmigration from sequences to images. First, raw signals are stacked row-by-row as shown in Figure~\ref{fig:activity image} (a). After being permuted in the first loop (Line 6-18 in Algorithm~\ref{alg:activity frames}), each tri-axis sensor data has a chance to be adjacent to each of the other sensor data as shown in Figure~\ref{fig:activity image} (b). For example, supposing $N_r = 9$, then the final $S_p$ is $[1,2,3,4,5,6,7,8,9,1,3,5,7,9,2,4,6,8,1,4,7,1,5,8,2,5,\\9,3,6,9,4,8,3,7,2,6,1]$. Since we still need to extract the relationships between each pair of single sensor signals, the second loop (Line 19-25 in Algorithm~\ref{alg:activity frames}) ensures that each single signal has a chance to be adjacent to each of the other signals as Figure~\ref{fig:activity image} (c) shows. So far we have extracted the relationships between each pair of single sensor signals.



\newcommand{\myindent}[1]{
\newline\makebox[#1cm]{}
}

\begin{algorithm}[!t]
\caption{Transformation from Sequences to Activity Frames}
\label{alg:activity frames}
\begin{algorithmic}[1]
\SetAlgoNoLine
\renewcommand{\algorithmicrequire}{\textbf{Input:}}
\renewcommand{\algorithmic}{\textbf{Hyper-parameters:}}
 \renewcommand{\algorithmicensure}{\textbf{Output:}}
 \REQUIRE Stacked raw signals. 
 Each row is a tri-axis data \myindent{0.3} of a accelerometer, gyroscope or a magnetometer\myindent{0.3} which can be denoted as $x, y, z$.
As shown in \myindent{0.3} Figure~\ref{fig:activity image} (a), each row has a sequence number.  \myindent{0.3} Here the number of rows $N_r$ = 9 as an example.
 \ENSURE  The activity frame $I_A$ which is a 2-D array
\STATE $i = 1;$

\STATE $j = i + 1;$

\STATE permutation sequence $S_p = [0];$

\STATE adjacent pair set $S_ap = \emptyset;$

\STATE activity frame $I_A$ = the first row of stacked signals

\WHILE{$i \neq j$}

\IF{$j > N_r$}
\STATE     {$j = 1$\;}
\ELSIF{$(i,j) \not\in S_ap$ and $(j,i) \not\in S_ap$}

\STATE add $(i, j)$ to $S_ap$;
 
\STATE  add $j$ to $S_p$;
 
\STATE  add the $j$-th row of input data to $I_A$;
 
\STATE  $i = j;$
 
\STATE  $j = i + 1;$

\ELSE
\STATE $j = j + 1$

\ENDIF
\ENDWHILE

\FOR{each row of $I_A$}
\IF{the sequence number of this row is odd}
\STATE this row is extended as $'x, y, z, x, y, z, x, y, z'$
\ELSE
\STATE this row is extended as $'x, y, z, y, z, x, z, x, y'$
\ENDIF

\ENDFOR
\RETURN $I_A$

\end{algorithmic}
\end{algorithm}

\subsection{Parallel Recurrent Model with Convolutional Attentions}

\begin{figure}[htbp] 
\centering
\includegraphics[width=3.2in]{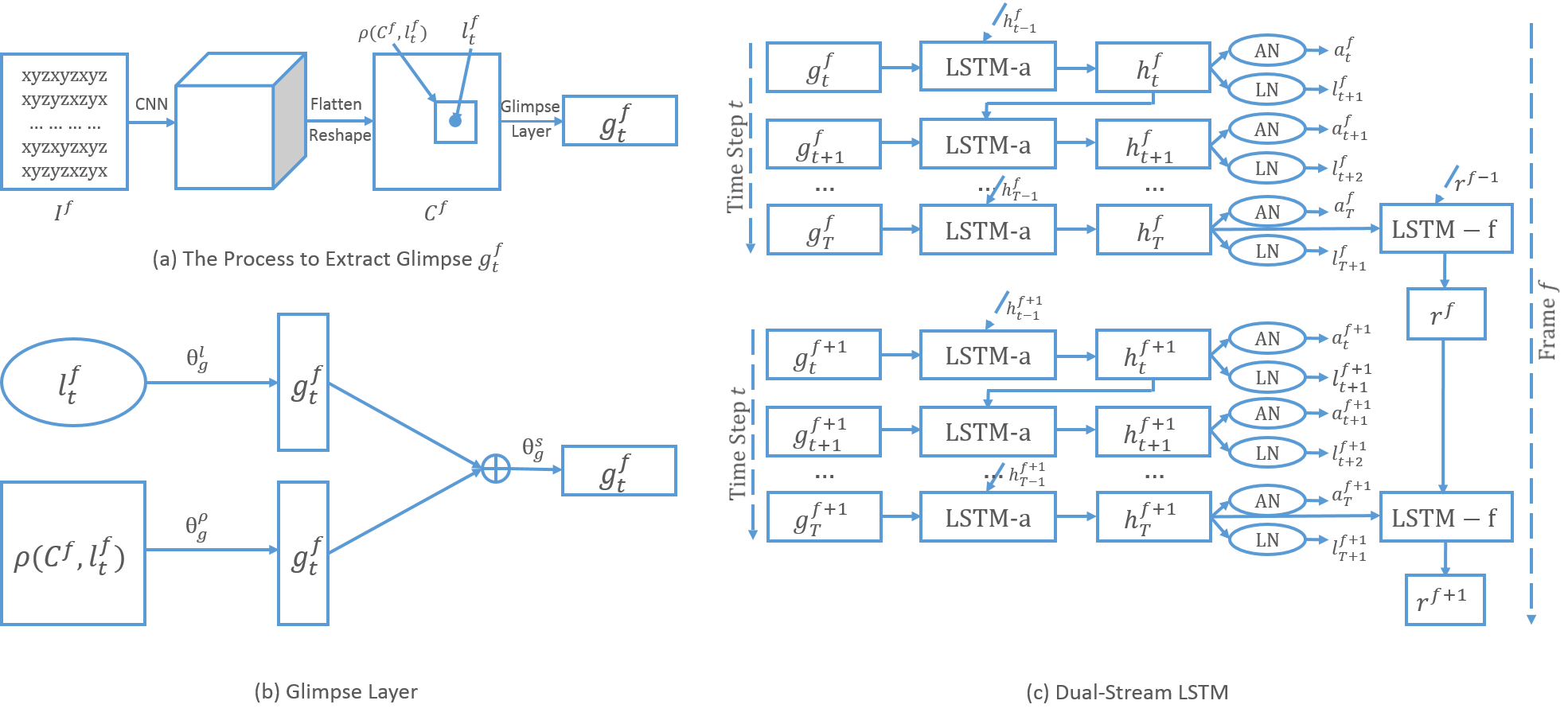} 
\caption{Flattened Model. (a) Extracting glimpse $g^f_t$ from the input activity frame, including a CNN, flattening and reshaping, and a glimpse layer. (b) The detailed description of the glimpse layer which combines the location $l^f_t$ and the retina region $\rho (C^f, l^f_t)$. (c) Parallel recurrent procedure containing attention based LSTM-$a$ and activity frame based LSTM-$f$.}
\label{fig:flattened model} 
\end{figure} 

We propose an interpretable parallel recurrent model with convolutional attentions
that incorporates both attention and temporal information to analyze the activity frames. Figure~\ref{fig:flattened model} shows the structure of this model, where the activity frame based LSTM recurrent model leverages the temporal information of sensor data and the attention based LSTM solves the HAR problem.

Since human body parts contribute differently in recognizing various activities, we need to guarantee that the system only focuses on the most relevant and contributing parts and data. Some previous works \cite{mnih2014recurrent,denil2012learning} leverages attention for image classification. However, analyzing activity sensor data in this work can be more challenging because sensor data lacks meaningful characteristics compared with image data. Therefore, we combine attention mechanism with CNN and RNN to automatically extract the most salient modality-specific features, further convert the information to higher-level representation and combine the spatial information with temporal information. Since the activity frames fully extract the relationships among all feature pairs, only a part of them is salient to each certain activity. Therefore, it is natural to introduce attention mechanisms facilitating to mine effective information and minimize the negative impacts of undesirable information. To the best of our knowledge, our method is the first one to leverage the attention model to tackle the activity recognition problems.
 
Figure~\ref{fig:flattened model} shows a flattened model, which better interprets the model. Our model is comprised of a convolutional network, a glimpse network, a recurrent attention unit, and a recurrent activity frame unit that we will introduce in the followings.

\subsubsection{\textbf{Convolutional Network}}

To derive an effective representation of features, we further transform activity frames into convolutional activity frames. Compared with a convolutional auto-encoder \cite{haque2016recurrent}, we prefer to train the model end-to-end and omit the pretraining process, as shown in Figure~\ref{fig:flattened model}. Each activity frame $I^f$ ($f$ denoted the $f_{th}$ frame) is transformed into a three-dimensional cube, the height of which depends on the number of channels of the convolutional network. The convolutional network has two convolutional layers that learn filters which activate when it detects some specific types of features at some spatial position in the input. The output is further processed by a ReLU layer and a max pooling layer.
The former applies the non-saturating activation function $relu(\nu) = max(\nu,0)$ to increase the nonlinear properties of both the decision function and the overall network without affecting the receptive fields of the convolution layer.
The latter partitions the input image into a set of non-overlapping rectangles and outputs the maximum for each such sub-region to omit the less important features.

To obtain new convolutional activity frames, the cubes are flattened and reshaped in the same size of original activity frames by a fully connected layer. After the convolutional layer, the input frame $I^f$ is encoded to be $C^f$.

 \subsubsection{\textbf{Glimpse Network}}
 \label{sec:Glimpse Network} 
 
The first part after the convolutional network is a glimpse network. The glimpse network not only avoids the system processing the whole data in the entirety at a time but also maximally eliminates the information loss. In our model, each frame will be "understood" within $T$ glimpses. For the transformed frame $C^f$, at each time step $t$, we simulate the process of how the human eyes work. Our model first extracts a retina region denoted by $\rho (C^f, l^f_t)$ from the input data at the location $l^f_t$ with a retina. The retina image encodes the region around $l^f_t$ with high resolution but uses a progressively lower resolution for points further from $l^f_t$. This has been proved an effective method to remove noises and avoid information loss in \cite{zontak2013separating}. 
 
In the human visual system, the retina image is converted into electric signals that are relayed to the brain via the optic nerves. Likewise, in our model, the retina image is converted into a glimpse $g^f_t$ as Figure~\ref{fig:flattened model} shows. The retina image $\rho (C^f, l^f_t)$ and the location $l^f_t$ are linearly transformed independently with two linear layers parameterized by $\theta _g^\rho$ and $\theta_g^l$, respectively. Next, the summation of these two parts is further transformed with another linear layer parameterized by $\theta _g^s$ and a rectified linear unit. The whole process can be summarized as the following equation:
\begin{align*}
g^f_t &= f_g(\rho (C^f, l^f_t), l^f_t;{\theta _g^\rho, \theta_g^l, \theta_g^s}) \\
&= relu(Linear(Linear(\rho (C^f, l^f_t)) + Linear(l^f_t)))
\end{align*}
where $Linear(\bullet)$ denotes a linear transformation. Therefore, $g^f_t$ contains information from both "what" ($\rho (C^f, l^f_t)$) and "where" ($l^f_t$).
 
\subsubsection{\textbf{Recurrent Attention Unit}}
 
We use the recurrent neural networks as the core to process data step by step within several glimpses and introduce an attention mechanism to ensure the system only focuses on the most relevant sensors/modals and the most contributing data. The glimpses at time steps of the attention based LSTM help visualize the contribution of sensors deployed at different body parts, thus achieving better interpretability of our model.

As Figure~\ref{fig:flattened model} shows, the basic structure of the recurrent attention unit is an LSTM-$a$ (attention based LSTM). At each time step $t$, the LSTM-$a$ receives the glimpse $g^f_t$ and the previous hidden state $h^f_{t-1}$ as the inputs parameterized by $\theta _h$. Meanwhile, it outputs the current hidden state $h^f_t$ according to the equation:

\begin{equation}
\label{eqn:h}
h^f_t = f_g(h^f_{t-1}, g^f_t;\theta _h)
\end{equation}

The recurrent attention model also contains two sub-networks: the location network and the action network. These two sub-networks receive the hidden state $h_t^f$ as the input to decide the next glimpse location $l^f_{t+1}$ and the current action $a^f_t$.
The current action not only determines the activity label $\hat{\textbf{y}}$ but also affects the environment in some cases while the location network outputs the location at time $t+1$ stochastically according to the location policy defined by a Gaussian distribution stochastic process, parameterized by the location network $f(h_t^f; \theta _t)$. 
As it decides the next region to "look at", the location network is the principal component of the recurrent attention unit.

\begin{equation}
\label{eqn:l}
l^f_{t+1} \sim P(\cdot \mid f_l(h_t^f; \theta _l))
\end{equation}

Similarly, the action network outputs the corresponding action at time $t$ and predicts the activity label given the hidden state $h_t^f$. The action $a_t^f$ obeys the distribution parameterized by $f(h_t^f; \theta _a)$. Owing to its prediction function, the network uses a softmax formulation:

\begin{equation}
\label{eqn:a}
a^f_t = f_a(h_t^f; \theta _a) = softmax(Linear(h_t^f))
\end{equation}

\subsubsection{\textbf{Recurrent Activity Frame Unit}}
Activity recognition heavily relies on the temporal information.
Therefore, besides the single activity frames used by the aforementioned process,
we additionally leverage activity frames via a recurrent activity frame unit.
As the hidden layer $h_t^f$ of the core LSTM-$a$ contributes to predicting the action $a_t^f$ and deciding the next glimpse location $l_{t+1}^f$.
For this reason, we believe the hidden state is discriminative enough to make the final prediction for the whole system.
In particular, we design an LSTM-$f$ (activity frame based LSTM) to combine the hidden states of all the frames at the last time step $T$ to predict the activity label and to preserve the efficiency.
Given the hidden state of the last frame, the hidden state of each frame $r^f = f_r(h_T^f, r^{f-1}; \theta _r)$, parameterized by $\theta _r$.

\subsection{Training and Optimization}

Our proposed model depends on the parameters of every components, including the glimpse network, the recurrent attention network, the two sub-networks, and the activity frame based LSTM, $\Theta = {\theta_g, \theta_h, \theta_a, \theta_l, \theta_r}$. Both the action network and the frame-based recurrent network are based on classification methods. Therefore, their parameters, $\theta_a$ and $\theta_r$, can be trained by optimizing the cross-entropy loss and the backpropagation. However, the location network should be able to select a sequence of salient regions from activity frames adaptively. Since this network is non-differentiable owing to its stochasticity and the problem can also be regarded as a control problem to settle the attention region at the next step, it can be trained by reinforcement methods to learn the optimal policies.

We simply introduce some definitions of reinforcement learning based on our case.
\begin{itemize}
\item Agent: the brain to make decisions, which is the location network in our case.
\item Environment: the unknown world that may affect the agent's decision or may be influenced by the agent.
\item Reward: the feedback from the environment to evaluate the action. In our case, for each frame, the model gives a prediction $\hat{\textbf{y}} = a_t$ and receives a reward $r_t$ as a feedback for the future correction of the prediction after each time step $t$. Suppose $T$ denotes the number of steps in our attention based LSTM. $r_t = 1$ if $\hat{\textbf{y}} = \textbf{y}$ after $T$ steps and $0$ otherwise. The target of the optimization is to maximize $R = \sum_{t=1}^{T} r_t$.
\item Policy: the projection from states to actions, denoted by $\pi(a\mid s) = P[A_t=a\mid S_t=s]$. To maximize the reward $R$, we learn an optimal policy $\pi(l_t,a_t|s_{1:t};\Theta)$ to map the attention sequence $s_{1:t}$ to a distribution over actions for the current time step, where the policy $\pi$ is decided by $\Theta$ of the recurrent attention model.

\end{itemize}

Based on the above discussion, we deploy a Partially Observable Markov Decision Process (POMDP) to solve the training and optimization problems, for which the true state of the environment is unobserved. Let $s_{1:t} = \textbf{x}_1, l_1, a_1;... \textbf{x}_t, l_t, a_t$ be the sequence of the input, location and action pairs. This sequence, called an attention sequence, shows the order of the regions our attention focuses on. 

To sum up, in our case, the location network is formulated as a random stochastic process (the Gaussian distribution) parameterized by $\Theta$. Each time after the location selection, the prediction $a$ is evaluated to back feed a reward for conducting the backpropagation training process. The process is also defined as policy gradient. Our goal is to maximize the simulated rewards using gradient.

Generally, for sample $x$ with its reward $f(x)$ and the probability $p(x)$, we have:
\begin{equation}
E_x[f(x)] = \sum_{x}p(x)f(x)
\end{equation}

so that the gradient can be calculated according to the REINFORCE rule \cite{williams1992simple}:

\begin{eqnarray}
\label{eqn:R rule}
\triangledown_\theta E_x[f(x)] &=& \triangledown_\theta\sum_{x}p(x)f(x) \nonumber \\
&=& \sum_{x}\triangledown_\theta p(x)f(x) \nonumber\\
&=& \sum_{x}p(x)\frac{\triangledown_\theta p(x)}{p(x)}f(x) \nonumber\\
&=& \sum_{x}p(x)\triangledown_\theta log p(x)f(x)      \nonumber \\
&=& E_x[f(x)\triangledown_\theta log p(x)]
\end{eqnarray}

In our case, given the reward $R$ and the attention sequence $s_{1:T}$, the reward function to be maximized is as follows:

\begin{equation}
\label{eqn:J}
J(\Theta) = \mathbb{E}_{p(s_{1:T};\Theta)}[\sum_{t=1}^{T} r_t] = \mathbb{E}_{p(s_{1:T};\Theta)}[R]
\end{equation}

By considering the training problem as a POMDP, a sample approximation to the gradient is calculated as follows:

\begin{equation}
\triangledown_\Theta J = \sum_{t=1}^{T} \mathbb{E}_{p(s_{1:T};\Theta)}[\triangledown_\Theta log \pi(\textbf{y}|s_{1:t};\Theta)R]
\end{equation}
where $i$ denotes the $i^{th}$ training sample, $\textbf{y}^{(i}$ is the correct label for the $i^{th}$ sample, and $\triangledown_\Theta log \pi(\textbf{y}^{(i)}|s_{1:t}^i;\Theta)$ is the gradient of LSTM-$a$ calculated by backpropagation.

We use Monte Carlo sampling which utilizes randomness to yield results that might be deterministic theoretically. Supposing $M$ is the number of Monte Carlo sampling copies, we duplicate the same convolutional activity frames for $M$ times and average them as the prediction results to overcome the randomness in the network, where the $M$ duplication generates $M$ subtly different results owing to the stochasticity, so we have:
\begin{equation}
\begin{aligned}
\nabla_\Theta J = \sum_{t=1}^{T}\mathbb{E}_{p(s_{1:t};\Theta)}[\nabla_\Theta log \pi(\textbf{y}|s_{1:t};\Theta)R]  \\
\thickapprox \frac{1}{M} \sum_{i=1}^{M} \sum_{t=1}^{T} \nabla_\Theta log \pi(\textbf{y}^{(i)}|s_{1:t}^i;\Theta)R^{(i)}
\end{aligned}
\end{equation}

Therefore, although the best attention sequences are unknown, our proposed model can learn the optimal policy in the light of the reward.

\section{Experiments}
In this section, we present the systematically validation of our proposed method via experiments on on two public datasets and one real-world dataset collected by ourselves. 

\subsection{Datasets and Experimental Settings}

We evaluate the proposed method on two public benchmarked activity recognition datasets, PAMAP2 dataset and MHEALTH dataset and the real-world dataset MARS which is collected by ourselves.
These public datasets are the latest available wearable sensor-based datasets with complete annotation and have been widely used in the activity recognition research community.

\textbf{PAMAP2} \cite{reiss2012introducing}: The PAMAP2 dataset contains 9 participants performing 12 daily living activities including both basic actions and sportive exercises. The activity sensory data is collected from 3 Inertial Measurement Units (IMUs) attached to three different positions, namely the dominant wrist, the chest and the dominant side’s ankle. Each IMU contains two 3-axis accelerometers, one 3-axis gyroscopes, one 3-axis magnetometers and one thermometer with the sampling rate of 100 Hz. 

\noindent\textbf{MHEALTH} \cite{banos2015design}: The MHEALTH dataset is also devised to benchmark methods of human activities recognition based on multimodal wearable sensor data. Three IMUs were respectively placed on the participants' chest, right wrist, and left ankle to record the acceleration ($ms^{-2}$), angular velocity (deg/s) and the magnetic field (local) data while they were performing 12 activities. The IMU on the chest also collected 2-lead ECG data (mV) to monitor the electrical activity of the heart. All sensing models are recorded at the frequency of 50 Hz.

\noindent\textbf{MARS} The MARS (Multimodal Activity Recognition with Sensing) dataset was collected while 8 participants (6 males, 2 females) were doing 5 basic activities (sitting, standing, walking, ascending stairs and descending stairs). Three IMU sensors, Phidget Spatial 3/3/3, were attached to the dominant wrist, the waist, and the dominant side's ankle, respectively, to collect the acceleration (gravitational acceleration $g$), angular velocity ($^{\circ}/s$), and magnetism ($nT$). 
All IMUs collected the data at the frequency of 70 Hz.

Similar to \cite{guo2016wearable}, the experiments conducted on the two public datasets perform background activity recognition task \cite{reiss2012introducing}. The activities are categorized into 6 classes: lying, sitting/standing, walking, running, cycling and other activities. To tackle the task and ensure the rigorousness, all experiments are performed by Leave-One-Subject-Out (LOSO) cross-validation which can also test the person independence during the evaluation.

\textbf{Model Implementation: }
In our experiments, we fix the size of glimpse to be $5\times5$ and deploy a CNN with two convolution layer, one ReLU layer and one max pooling layer.
The dimensionality of $g\rho^f_t, gl^f_t$, $g^f_t$,  $\theta_a$ and $M$ are fixed to 128, 128, 220, 100 and 20 respectively. The location network is defined by a dual-component Gaussian with a variance of 0.22. The proposed method has two recurrent networks. One is the attention based LSTM with the cell size of 100. The number of time steps is 40, which defines the number of glimpses. The other one is the frame-based recurrent network with LSTM size of 1000 and time steps set to 5.

The proposed model is trained on a Nvidia Titan X Pascal GPU from scratch. The stochastic gradient descent with Adam update rule is used to minimize the cross-entropy loss function. The network parameters are optimized with a linearly annealed learning rate from 0.01 to $10^{-5}$. To ensure the rigorousness, the experiments are performed by Leave-One-Subject-Out (LOSO) cross-validation.

\subsection{Accuracy Comparison and Performance Analysis}

To evaluate the performance of the proposed approach, our model, we conduct extensive experiments to compare its performance with the state-of-the-art methods on PAMAP2 and MHEALTH. We elaborately select other four state-of-the-art and multimodal feature-based approaches (MARCEL \cite{guo2016wearable}, FEM \cite{lara2012centinela}, CEM \cite{guo2014activity} and MKL \cite{althloothi2014human}) and five baseline methods (Support Vector Machine (SVM), Random Forest(RF), K-Nearest Neighbors(KNN), Decision Tree(DT) and Single Neural Networks) to show the competitive power of the proposed method. To ensure fair comparison, the best parameters test, our model, is used on both datasets; the best trade-off parameter ($\lambda = 0.7$) is deployed for MARCEL; time-domain features including mean, variance, standard deviation, median and frequency-domain features including entropy and spectral entropy are utilized for FEM; each modality feature group are defined an independent kernel for MKL; and for other baseline methods, all modality features are deployed. All parameters adopted are in reference to the parameters suggested in literature. The results in Table~\ref{tab:system_comparison} show the proposed our model outperforms all the state-of-the-art methods and the baseline methods.
\begin{table}[ht]
\centering
\caption{Comparison among our model and four state-of-the-art methods and five baseline methods. For PAMAP2 dataset, accelerometer, gyroscope and magnetism are utilized. For MHEALTH dataset, ECG data are considered additionally.}
\label{tab:system_comparison}
\begin{tabular}{ccc}
\hline
\multirow{2}{*}{\textbf{Method}} & \multicolumn{2}{c}{\textbf{Accuracy}}\\
\cline{2-3}
&\textbf{PAMAP2}&\textbf{MHEALTH}\\
\hline
MARCEL \cite{guo2016wearable} & 82.8 & 92.3\\
FEM+SVM \cite{lara2012centinela} & 76.4 & 70.7 \\
CEM \cite{guo2014activity} & 81 & 74.8 \\
FEM+MKL \cite{lara2012centinela, althloothi2014human} & 81.6 & 90.6\\
\hline
SVM & 59.3 & 68.7 \\
Random Forest & 64.7 & 82.5 \\
KNN & 70.3 & 86.1 \\
Decision Tree & 57.8 & 78.7 \\
Single Neural Networks & 72.0 & 89.1 \\
\hline
\textbf{Our Model} & \textbf{83.4}& \textbf{94.0 }\\
\hline

\end{tabular}
\end{table}
\begin{figure}[htbp] 
\centering
\includegraphics[width=3.2in]{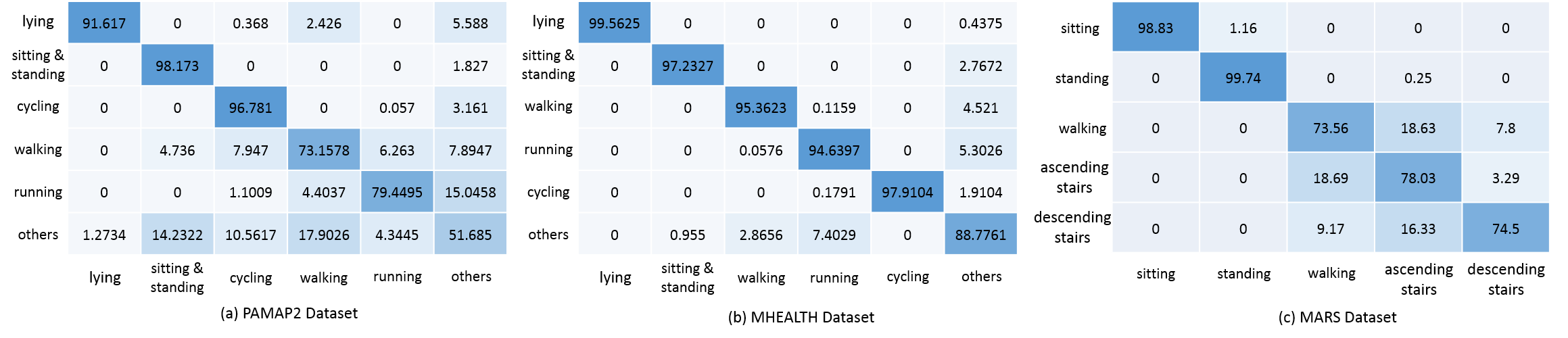} 
\caption{The confusion matrices of our model for background activity recognition on three datasets}
\label{fig:confusion_matrix_6_classes} 
\end{figure} 

To further explain the accuracy of our model on each specific activity, Figure~\ref{fig:confusion_matrix_6_classes} (a) and (b) show the confusion matrices on both public datasets performing the background activity recognition task.

The results show the proposed approach performs well for most activities such as lying, sitting and standing, and cycling.


\subsection{Effect of Activity Frames}
We prove the effectiveness of our model on original features. To adapt features to the proposed model, multimodal features are stacked to form original frames, as Figure~\ref{fig:activity image}(a) shows. Table~\ref{tab:feature extraction capability} presents feature extraction capability of activity frames, which shows that the proposed model based on the original frames outperforms most of the state-of-the-art methods (listed in Table~\ref{tab:system_comparison}) even without activity frames. But utilizing the activity frames can improve the performance of original model due to the availability of the full relationship among features provided by activity frames.

\begin{table}[ht]
\centering
\caption{Feature Extraction Capability of Activity Frames}
\label{tab:feature extraction capability}
\begin{tabular}{cccc}
\hline
& PAMAP2 & MHEALTH & MARS\\ \hline
Original Frames & 81.35 & 92.20 & 77.25\\
Activity Frames & \textbf{83.42} & \textbf{94.04} & \textbf{85.28}\\ \hline

\end{tabular}
\end{table}




\subsection{Model Interpretability}

\begin{figure}[htbp] 
\centering
\includegraphics[width=3.4in]{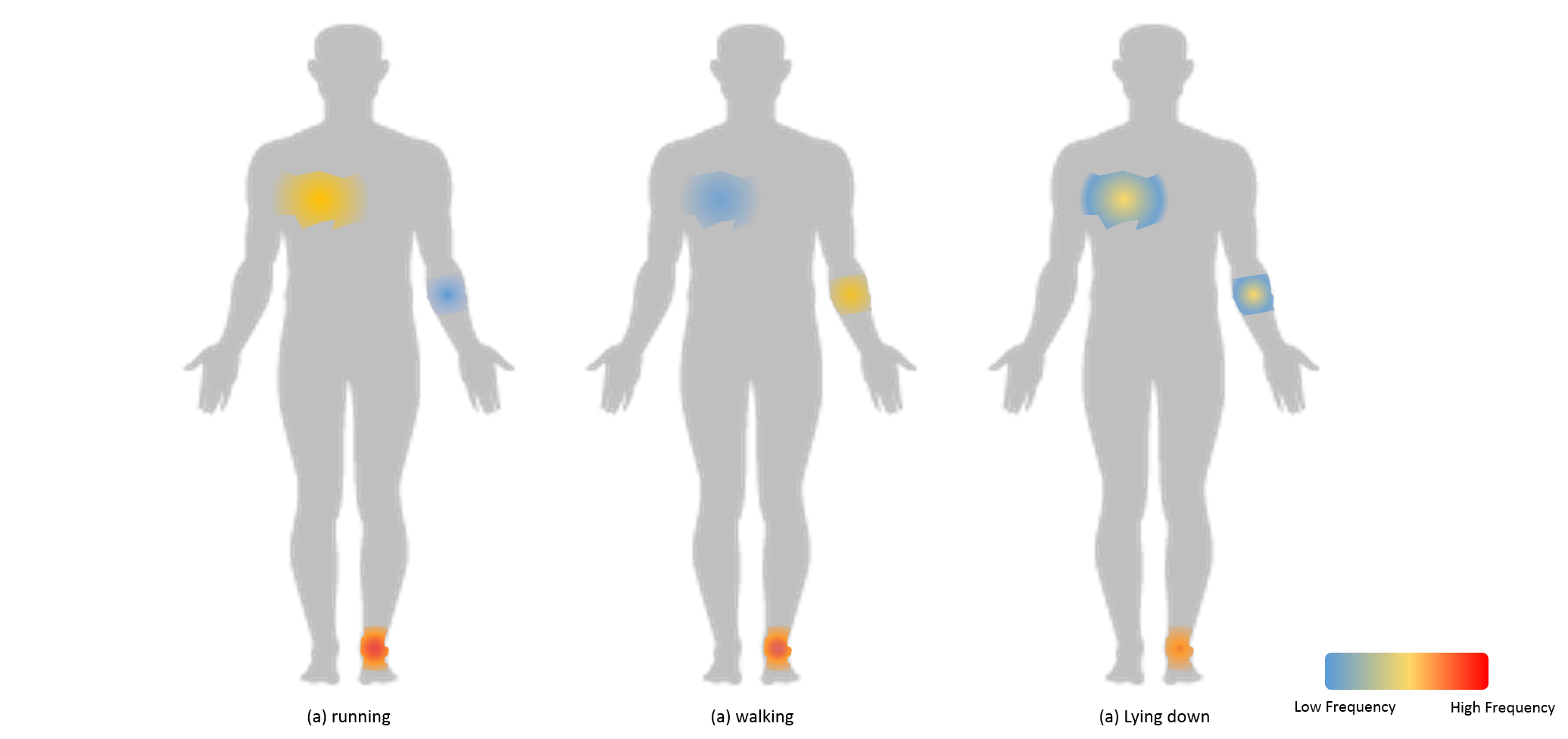} 
\caption{Glimpse Heatmap}
\label{fig: Glimpse Heatmap} 
\vspace{-0.5cm}
\end{figure}

One of the merits of our method is its interpretability. For wearable sensor-based activity recognition, subjects usually wear more than one sensors on their dominant body parts like arms, chest, and ankles, each sensor with multimodal. Attention mechanisms provide a superiority that it feeds the glimpse location back at each time step. Owing to the particularity of the activity frames, the attention model in our scenario not only provides the specific body parts it focuses on  but also highlights the most contributing sensors and modals to diverse activities. In this section, we only present the experimental results of running, walking and lying down on MHEALTH dataset for simplicity. The available sensors on MHEALTH include ECG, chest accelerometer, ankle accelerometer, ankle gyroscope, ankle magnetometer, arm accelerometer, arm gyroscope and arm magnetometer. Figure~\ref{fig: Glimpse Heatmap} shows the glimpse heatmap for all sensors. Taking running as an example, we can observe ankle as the most active part of running. The chest also contributes a lot while arm involves the least. To further demonstrate the involvement of all sensory modal data, Table~\ref{Tab: Modals Involvements on MHEALTH Dataset} concludes the percentage of our model "looking at" different modals for the latest 120 times (out of 200 times). It shows that for running, the most salient modal is ankle acceleration, which accounts for 30.55\%. ECG and ankle angular velocity are also significant. The experimental results totally conform to the reality that while running, the most active body parts should be legs and ankles. Another self-evident truth is that in our experiments, one modality that can easily distinguish strenuous exercise like running from others is ECG. Also, since the model still "looks at" other modals for several times, it is able to better corroborate the claim that our model minimizes information loss.

\begin{table*}[ht]
\centering
\caption{Modals Involvements on MHEALTH Dataset (\%). (Acc, Gyro, Magn denote Acceleration, Angular Velocity and Magnetism, respectively.)}
\label{Tab: Modals Involvements on MHEALTH Dataset}
\begin{tabular}{ccccccccc}
\hline
activity&ECG & $Acc_{chest}$ & $Acc_{ankle}$ & $Gyro_{ankle}$ & $Magn_{ankle}$ & $Acc_{arm}$ & $Gyro_{arm}$ & $Magn_{arm}$\\ \hline
running & \textbf{21.98} & 10.22 & \textbf{30.55} & \textbf{15.86} & 4.30 & 6.49 & 5.03 & 5.57\\ 
walking & 7.23 & 11.05 & \textbf{18.78} & \textbf{19.26} & 8.72 & \textbf{19.66} & 9.46 & 5.83\\
lying down & 6.58 & 13.45 & \textbf{16.34} & 10.23 & \textbf{17.92} & 10.29 & 10.72 & \textbf{14.47} \\

\hline
\end{tabular}
\end{table*}

\subsection{Hyper-Parameter Study}
\begin{figure}[htbp] 
\centering
\includegraphics[width=3.2in]{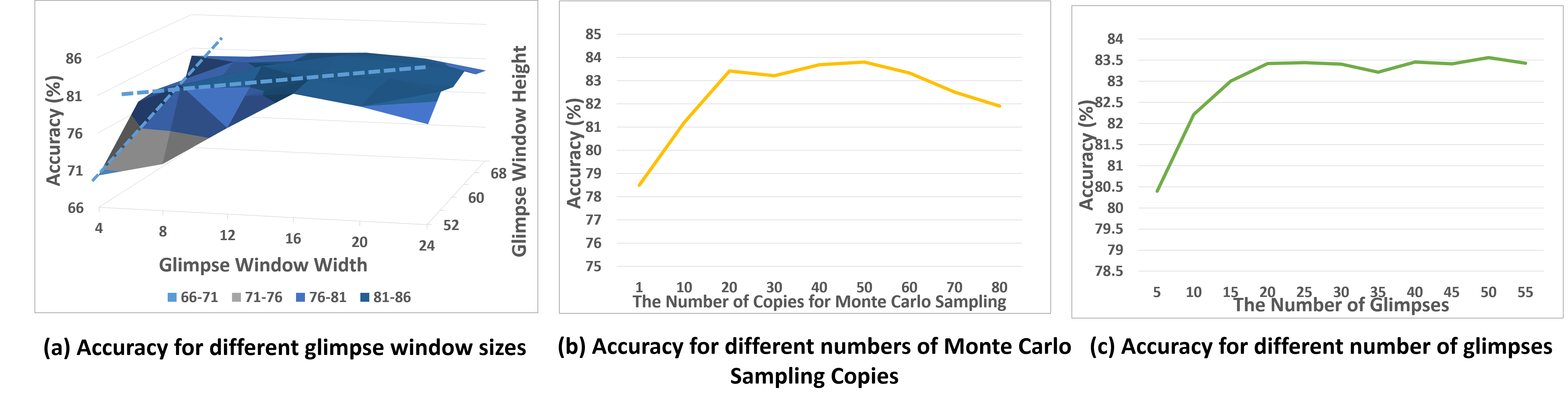} 
\caption{Experimental Results for Hyper-Parameter Tuning on PAMAP2}
\label{fig:parameter_tuning_pamap} 
\end{figure} 

\begin{figure}[htbp] 
\centering
\includegraphics[width=3.2in]{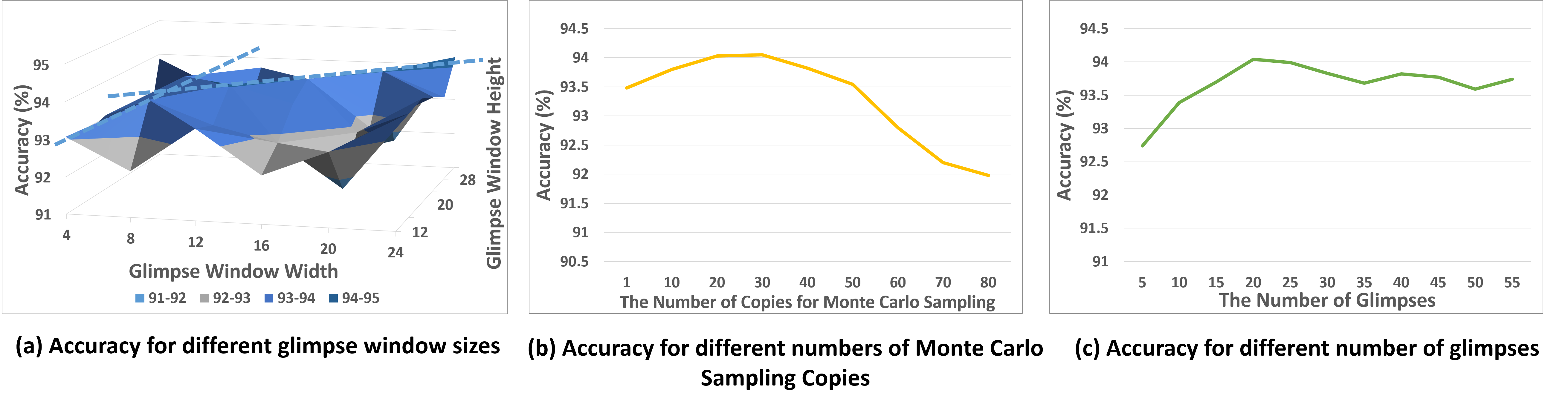} 
\caption{Experimental Results for Hyper-Parameter Tuning on MHEALTH}
\label{fig:parameter_tuning_mhealth} 
\end{figure} 

\begin{figure}[htbp] 
\centering
\includegraphics[width=3.2in]{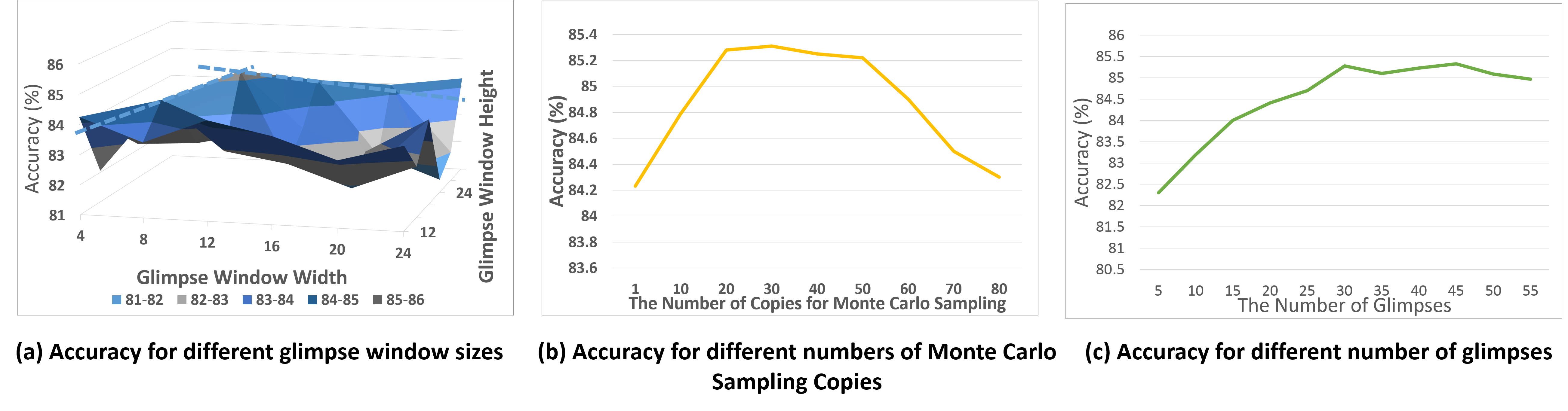} 
\caption{Experimental Results for Hyper-Parameter Tuning on MARS}
\label{fig:parameter_tuning_mars} 
\end{figure} 

In this section, we mainly analyze three most contributing hyper-parameters to which the model is more sensitive in our experiments, namely the size of glimpse windows (width and height), the number of copies for Monte Carlo sampling and the number of glimpses. For the other hyper-parameters, we just use fixed empirical values as suggested in the previous subsection. The variation trend is shown as Figure~\ref{fig:parameter_tuning_pamap}, Figure~\ref{fig:parameter_tuning_mhealth} and Figure~\ref{fig:parameter_tuning_mars}.

Taking Figure~\ref{fig:parameter_tuning_pamap} as an example, firstly, we tune the width and the height of glimpse windows to figure out their relationship as shown in Figure~\ref{fig:parameter_tuning_pamap} (a).
Specifically, there are 13 3-axis vectors to present the temperature, accelerometer, gyroscope and magnetism in our experiments. After Algorithm~\ref{alg:activity frames}, several $78$x$9$ activity frames are generated. Figure~\ref{fig:parameter_tuning_pamap} (a) shows that the accuracy achieves the best when the glimpse window size is $64$x$16$ and there is an obvious "ridge" along which the whole figure is almost symmetric. All the points on the symmetric line are in a ratio of $4:1$.
This suggests that the approach favors a fixed ratio of the two dimensions of the glimpse window, in spite that we use the ratio of the activity frame size of $78:9$. Also, we can see that Figure~\ref{fig:parameter_tuning_mhealth} (a) and Figure~\ref{fig:parameter_tuning_mars} (a) both show the "ridge" while their optional glimpse window sizes are different because of different sizes of activity frames.

Figure~\ref{fig:parameter_tuning_pamap} (b) and (c) show the experimental results of our studies on the effect of other two hyper-parameters, the number of copies for Monte Carlo sampling and the number of glimpses.
In particular, Figure~\ref{fig:parameter_tuning_pamap} (c) presents a trend that the accuracy increases remarkably at first and remains stable after getting a turning point. However, for the Monte Carlo sampling, too low or too high values lead to worse performance, as Figure~\ref{fig:parameter_tuning_pamap} (b) shows. And we can notice that the variation trends in Figure~\ref{fig:parameter_tuning_mhealth} and Figure~\ref{fig:parameter_tuning_mars} enjoy the same patterns. 

\section{Conclusion}

This paper proposes an innovative HAR approach, our model, which includes (a) a novel form of multimodal sensor features, activity frames to fully extract relations between each pair of sensors and modality data and (b) an interpretable parallel recurrent model with convolutional attentions to combine recurrent attention and recurrent activity frames. The experiments show our method outperforms the state-of-the-art methods. Also, the method enjoys great interpretability in spite of the non-explanation of neural networks.






\bibliographystyle{IEEEtran}
\bibliography{IJCNN}

\end{document}